\documentclass{aastex}          
\usepackage{spr-astr-addons}    
\usepackage{epsfig}
\usepackage{graphicx}
\usepackage{color}

\begin{document}
%
\title{Phase mixing of standing Alfv\'{e}n waves with shear flows in solar spicules}

\shorttitle{Phase mixing of Alfv\'{e}n waves in solar spicules}
\shortauthors{Ebadi et al.}

\author{H.~Ebadi\altaffilmark{1}}
\affil{Astrophysics Department, Physics Faculty,
University of Tabriz, Tabriz, Iran\\
e-mail: \textcolor{blue}{hosseinebadi@tabrizu.ac.ir}}
\and
\author{M.~Hosseinpour}
\affil{Plasma Physics Department, Physics Faculty,
University of Tabriz, Tabriz, Iran}

\altaffiltext{1}{Research Institute for Astronomy and Astrophysics of Maragha,
Maragha 55134-441, Iran.}

\begin{abstract}

Alfv\'{e}nic waves are thought to play an important role in coronal heating and solar wind acceleration.
Here we investigate the dissipation of such waves due to phase mixing at the presence of
shear flow and field in the stratified atmosphere of solar spicules.
The initial flow is assumed to be directed along spicule axis and to vary
linearly in the x direction and the equilibrium magnetic field is taken 2-dimensional and divergence-free.
It is determined that the shear flow and field can fasten the damping of standing Alfv\'{e}n waves. In spite of
propagating Alfv\'{e}n waves, standing Alfv\'{e}n waves in Solar spicules dissipate in a few periods.
As height increases, the perturbed velocity amplitude does increase in contrast to the behavior of
perturbed magnetic field. Moreover, it should be emphasized that the stratification due to gravity, shear flow
and field are the facts that should be considered in MHD models in spicules.

\end{abstract}


\section{Introduction}
\label{sec:intro}
Phase mixing has been proposed as a mechanism of efficiently dissipating Alfv\'{e}n waves in the solar corona by \citet{Hey1983}.
\citet{Karami2009} calculated numerically the damping times of standing Alfv\'{e}n waves
in the presence of viscosity and resistivity in coronal loops. They concluded
that the exponential damping law obtained by \citet{Hey1983} in time is valid for the Lundquist numbers higher than~$10^{7}$.
\citet{De1999, De2000} studied the effect of stratification and diverging background magnetic field on phase mixing, and found that
the wavelength of an Alfv\'{e}n wave is shortened as it propagates outwards which enhances the generation of gradients.
They concluded that the convection of wave energy into heating the plasma occurs at lower heights than in a uniform model.
Moreover, the combined effect of stratification and diverging background magnetic field depends on
the geometry of configuration.
\citet{Ruderman2007} showed that the enhanced phase mixing mechanism can dissipate Alfv\'{e}n waves at heights less than half.
Moreover, it can occur in divergent and stratified coronal structures only when the ratio of magnetic and density scale heights is
lower than 2.\\
\citet{kaghashvili1999} studied the effect of inhomogeneous flow on converting the Alfv\'{e}n waves into other types of MHD waves
that can dissipated efficiently. It is found that in the divergent geometry of magnetic fields, this mechanism becomes essential
in the rapidly expanding regions where the velocity shear flow is relatively large. \citet{Saleem2007} studied drift modes driven
by shear plasma flow in spicules and concluded that the density inhomogeneity and the shear in the plasma flow should be expected
as facts in their atmosphere. These waves propagating on spicules can be efficiently dissipated in the regions
with sheared magnetic field in such a way that can heat the solar corona.\\
Spicules appear as grass-like, thin and elongated structures in images of the solar lower atmosphere \citep{Tem2009, ster2000}.
\citet{Kukh2006, Tem2007} by analyzing the height series of $H\alpha$ spectra concluded
that transverse oscillations can be caused by propagating kink waves in spicules.
\citet{De2007,Okamoto2011} by making use of \emph{Hinode} observations interpreted that transverse oscillations of spicule axis
 are signatures of Alfv\'{e}n waves, and concluded that standing Alfv\'{e}n waves occurred at the
 rate of about $20\%$.
 More recently \citet{Ebadi2012a} based on \emph{Hinode}/SOT observations concluded that spicule axis oscillations
 are closed to the standing pattern as they do not see any upward or downward propagation.
 \citet{Ebadi2012b} performed the phase mixing model of propagating Alfv\'{e}n waves in spicule conditions
 and concluded that the damping times are much longer than spicule lifetimes.
 These results motivated us to study the phase mixing of standing Alfv\'{e}n waves in a stratified
 atmosphere in the presence of shear flow and field. To do so, the section $2$ gives the basic equations and the theoretical model. In section $3$ the numerical
 results are presented and discussed, and a brief summary is followed in section $4$.

\section{Theoretical modeling}
\label{sec:theory}
In the present work we keep the effects of stratification due to
gravity in $2$D x-z plane in the presence of shear flow and shear field.
The phase mixing and dissipation of standing
Alfv\'{e}n waves in a region with non-uniform Alfv\'{e}n velocity are
studied. MHD equations, governing the plasma dynamics are as follows:
\begin{equation}
\label{eq:cont} \frac{\partial \rho}{\partial t}+\nabla\cdot(\rho
\mathbf{v}) = 0,
\end{equation}
\begin{equation}
\label{eq:momentum} \rho\frac{\partial \mathbf{v}}{\partial t}+
\rho(\mathbf{v} \cdot \nabla)\mathbf{v} = -\nabla p + \rho
\mathbf{g}+ \frac{1}{\mu}(\nabla \times \mathbf{B})\times
\mathbf{B}+ \rho\nu\nabla^{2}\mathbf{v},
\end{equation}

\begin{equation}
\label{eq:induction} \frac{\partial \mathbf{B}}{\partial t} = \nabla
\times(\mathbf{v} \times \mathbf{B})+ \eta\nabla^{2}\mathbf{B},
\end{equation}

\begin{equation}
\label{eq:state} p = \frac{\rho RT}{\mu},
\end{equation}
\begin{equation}
\label{eq:divB} \nabla\cdot \mathbf{B} = 0,
\end{equation}
where $\nu$ and $\eta$ are constant viscosity and resistivity coefficients,
and other quantities have their usual meanings. In
particular, typical values for $\eta$ in the solar chromosphere and corona are
$8\times10^{8}T^{-3/2}$m$^{2}$s$^{-1}$ and $10^{9}T^{-3/2}$ m$^{2}$s$^{-1}$, respectively. The
value of $\rho\nu$ for a fully ionized H plasma is $2.2\times10^{-17}T^{5/2}$
kg m$^{-1}$ s$^{-1}$ \citep{Priest1982}.
We assume that the spicules are highly dynamic with speeds that are
significant fractions of the Alfv\'{e}n speed. The perturbations are assumed independent of y,
with a polarization in \^{y} direction, i.e.:
\begin{eqnarray}
\label{eq:perv}
  \textbf{v} &=& v_{0}(x) \hat{k} + v_{y}(x,z,t) \hat{j} \nonumber\\
  \textbf{B} &=& B_{0x}(x,z) \hat{i}+ B_{0z}(x,z) \hat{k} + b_{y}(x,z,t) \hat{j}.
\end{eqnarray}

The background flow is assumed to vary linearly in the x-direction as \citep{Rogava1998}:
\begin{equation}
\label{eq:shear flow}
\textbf{ v}_{0} = v_{0}(x)~\hat{k} = (x-1)v_{0}~\hat{k},
\end{equation}
and the equilibrium sheared magnetic filed is two-dimensional and divergence-free as \citep{Ruderman2007,Tem2010}:

\begin{eqnarray}
\label{eq:shear field}
 B_{0x}(x,z) &=& B_{0}e^{-k_{B}z} \sin[k_{B}(x-1)] \nonumber\\
 B_{0z}(x,z) &=& B_{0}e^{-k_{B}z} \cos[k_{B}(x-1)] ,
\end{eqnarray}

Therefore, the pressure gradient is balanced by the gravity force, which is assumed
to be \textbf{g}=-g $\hat{k}$ via this equation:

\begin{equation}
\label{eq:balance}
 -\nabla p_{0} + \rho_{0} \textbf{g}=0,
\end{equation}

and the pressure in an equilibrium state is:
\begin{equation}
\label{eq:press}
 p_{0}= p_{0}(x)e^{-z/H}.
\end{equation}

Moreover, the equilibrium density profile is given the form:

\begin{equation}
\label{eq:density}
 \rho_{0}= \rho_{0}(x)e^{-z/H},
\end{equation}
with
\begin{equation}
\label{eq:scale}
 H= \frac{RT}{\mu g},
\end{equation}
where $H$ is the pressure scale height.
Taking into account these assumptions, the linearized dimensionless form of Eqs.~\ref{eq:cont},
~\ref{eq:momentum}, and ~\ref{eq:induction} yield:

\begin{eqnarray}
\label{eq:velo}
  \frac{\partial v_{y}}{\partial t} &=& V^{2}_{A}(x,z)\left[B_{0x}(x,z)\frac{\partial b_{y}}{\partial x}+B_{0z}(x,z)\frac{\partial b_{y}}{\partial z}\right] \nonumber\\
   &  & -v_{0}(x)\frac{\partial v_{y}}{\partial z} + \nu\nabla^{2}v_{y},
\end{eqnarray}

and

\begin{eqnarray}
\label{eq:mag}
  \frac{\partial b_{y}}{\partial t} &=& \left[B_{0x}(x,z)\frac{\partial v_{y}}{\partial x}+B_{0z}(x,z)\frac{\partial v_{y}}{\partial z}\right] \nonumber\\
   &  & -v_{0}(x)\frac{\partial b_{y}}{\partial z} + \eta\nabla^{2}b_{y},
\end{eqnarray}

where velocities, magnetic field, time and space coordinates are normalized to V$_{A0}\equiv B_{\rm 0}/\sqrt{\mu \rho_{\rm 0}}$
(with $\rho_{\rm 0}$ as the plasma density at $z=0$),
$B_{\rm 0}$, $\tau$ (Alfv\'{e}n time), $a$ (spicule radius), respectively. Also the resistivity and viscosity coefficients
are normalized to $a^{2}/\tau$.
The first and second terms in the right hand side of Eqs.~\ref{eq:velo}, and~\ref{eq:mag} representing effects of shear fields and flows, respectively.
$V_{\rm A}(x,z)$ is the Alfv\'{e}n velocity, which for
a phase mixed and stratified atmosphere due to gravity is
assumed to be \citep{De1999,Karami2009}:

\begin{equation}
\label{eq:av}
 V_{A}(x,z)=V_{A0}e^{z/2H}[2+\tanh[\alpha(x-1)]].
\end{equation}
Here, parameter $\alpha$ controls the size of inhomogeneity across the magnetic field. \\
The set of Eqs.~\ref{eq:velo}, and~\ref{eq:mag} should be solved under these
initial conditions:

\begin{eqnarray}
\label{eq:icv}
  v_{y}(x,z,t=0) &=& V_{A0}\exp \left [-\frac{1}{2}(\frac{x-1}{d})^{2}\right]\sin(kz)e^{z/4H} \nonumber\\
  b_{y}(x,z,t=0) &=& A \sin(\pi x)\sin(\pi z) ,
\end{eqnarray}
where d is the width of the initial packet and $A=10^{-7}$.
\begin{figure}
\centering
\includegraphics[width=8cm]{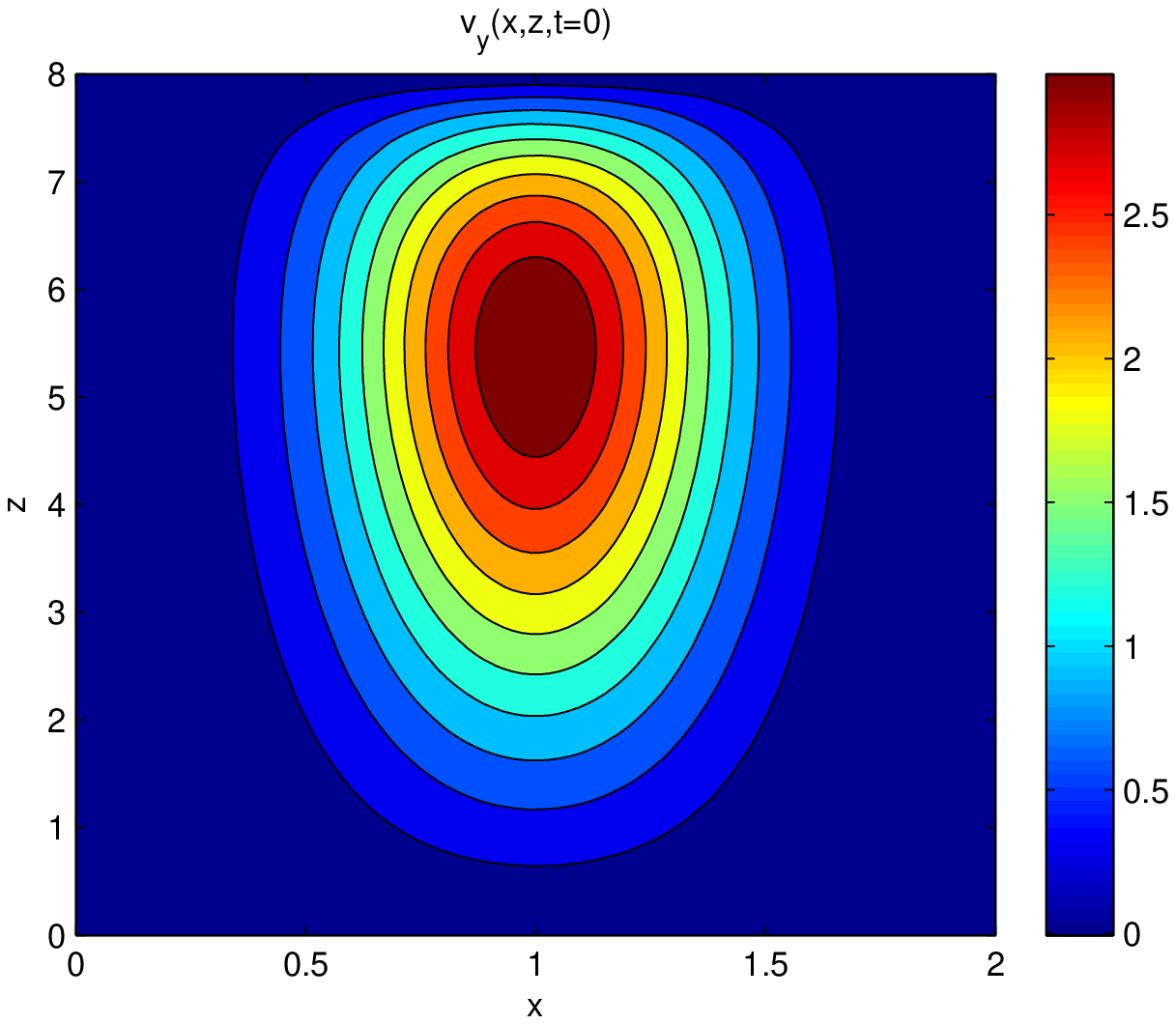}
\includegraphics[width=8cm]{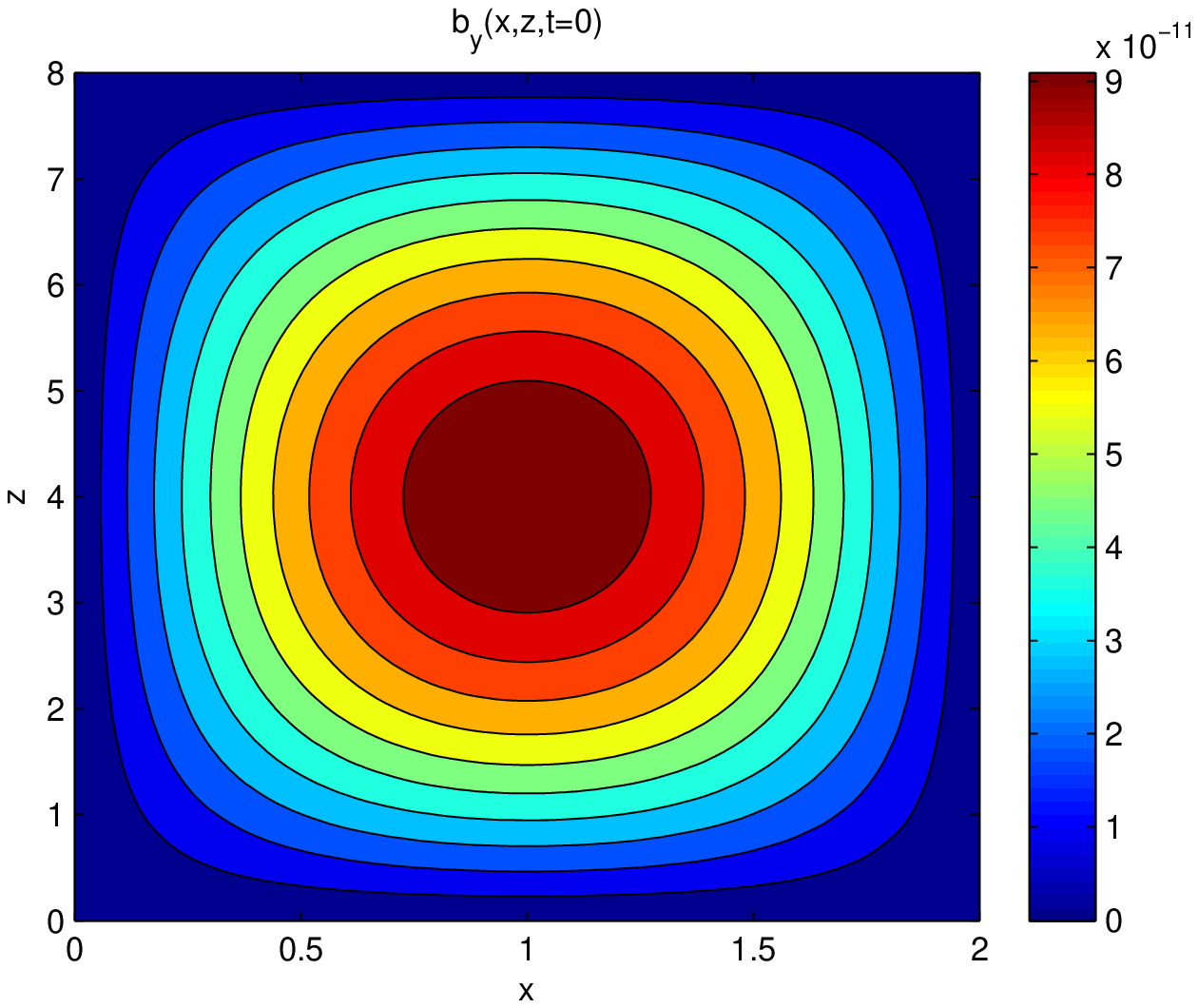}
\caption{(Color online) The initial wave packet and magnetic field in $x-z$ space are presented from top to bottom, respectively. \label{fig1}}
\end{figure}
Figure~\ref{fig1} is the plot of initial wave packet and magnetic field given by
Eq.~\ref{eq:icv} for $d=0.3a$ ($a$ is the spicule radius). Also, the boundary conditions are taken to be:
\begin{eqnarray}
\label{eq:icx}
  v_{y}(x=0,z,t)=v_{y}(x=2,z,t)=0 \nonumber\\
  b_{y}(x=0,z,t)=b_{y}(x=2,z,t)=0,
\end{eqnarray}
and
\begin{eqnarray}
\label{eq:icz}
  v_{y}(x,z=0,t)=v_{y}(x,z=8,t)=0 \nonumber\\
  b_{y}(x,z=0,t)=b_{y}(x,z=8,t)=0.
\end{eqnarray}

Boundary conditions in Eq.~\ref{eq:icz} imply that we are dealing with standing waves.

\section{Numerical results and discussion}

\begin{figure}
\centering
\includegraphics[width=8cm]{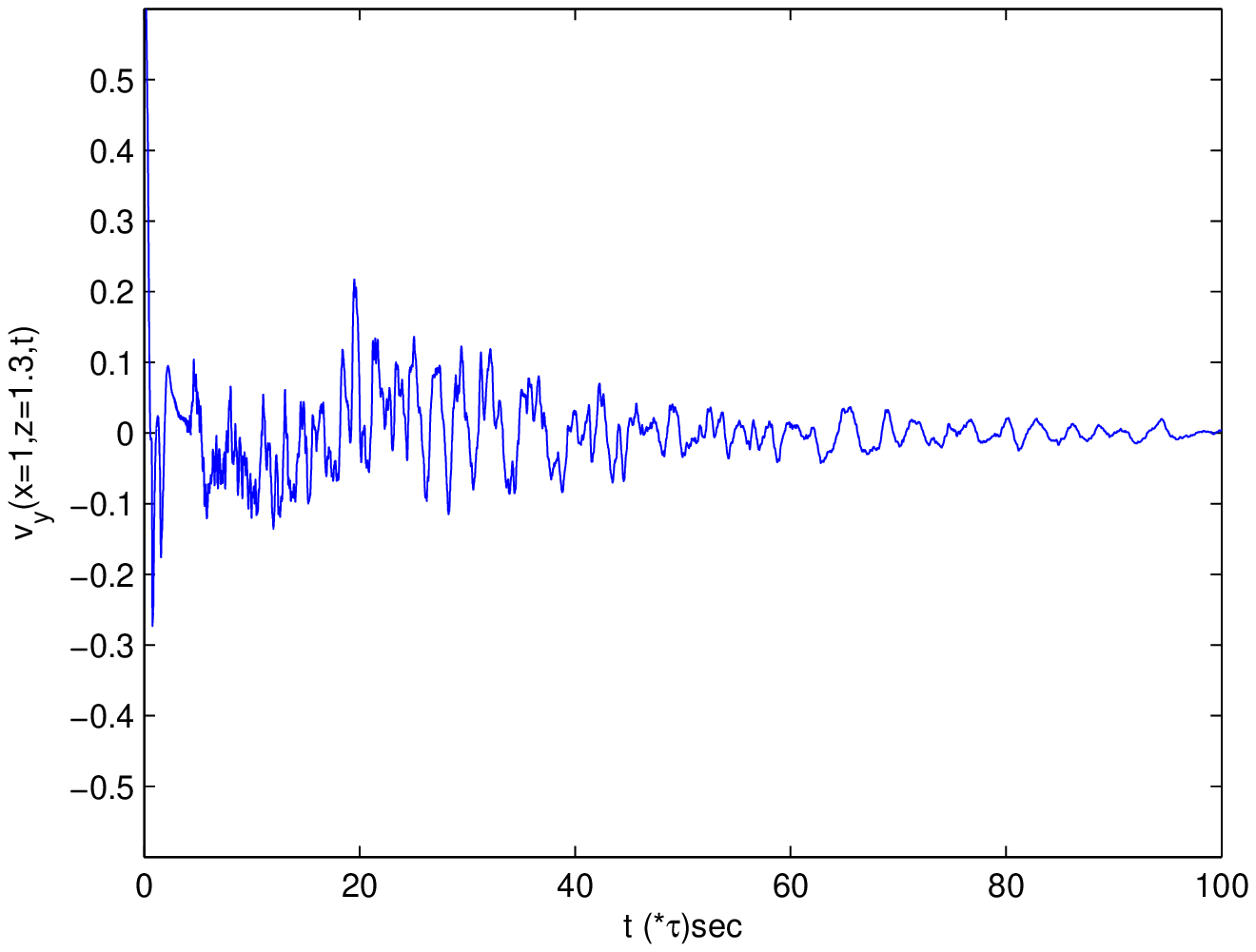}
\includegraphics[width=8cm]{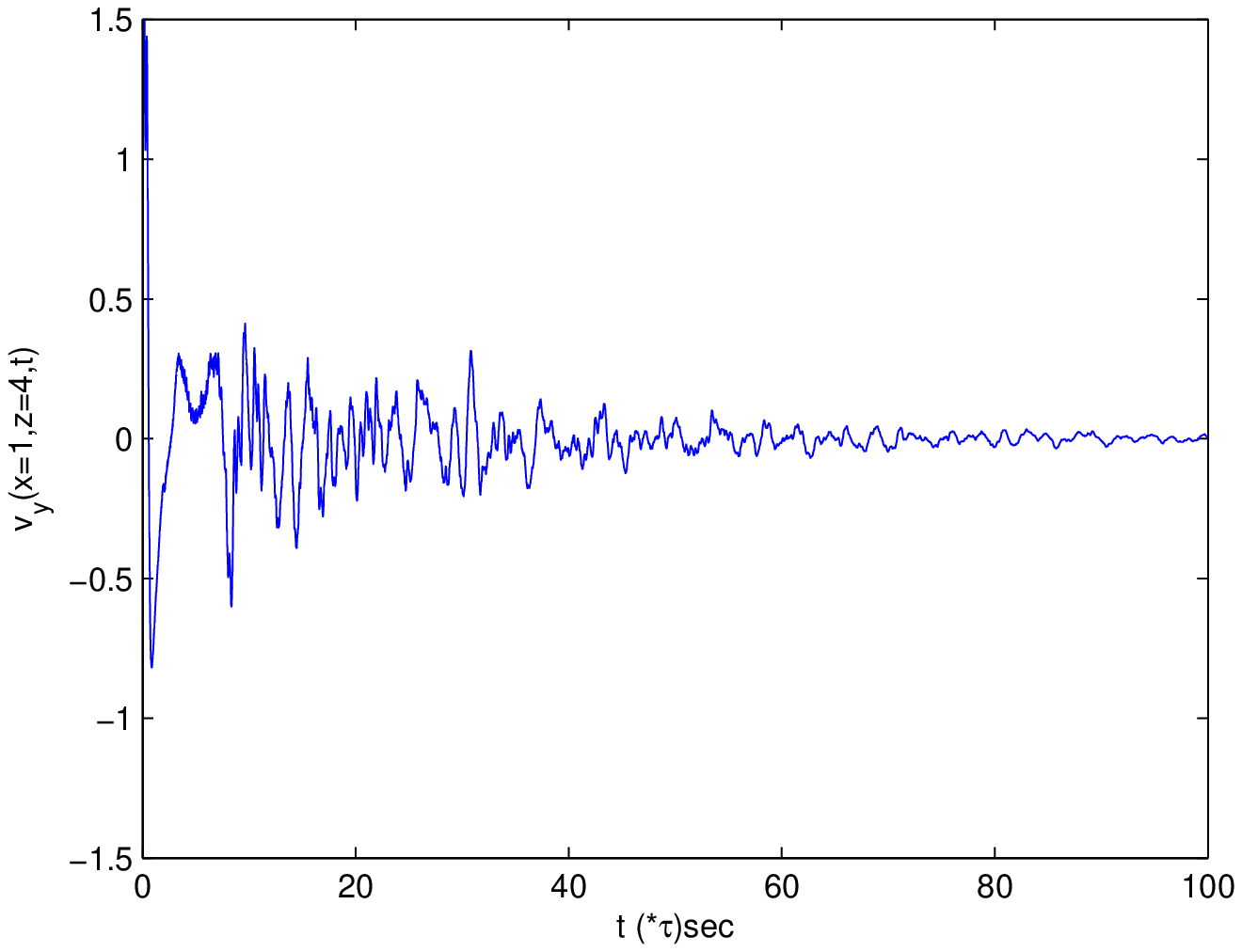}
\includegraphics[width=8cm]{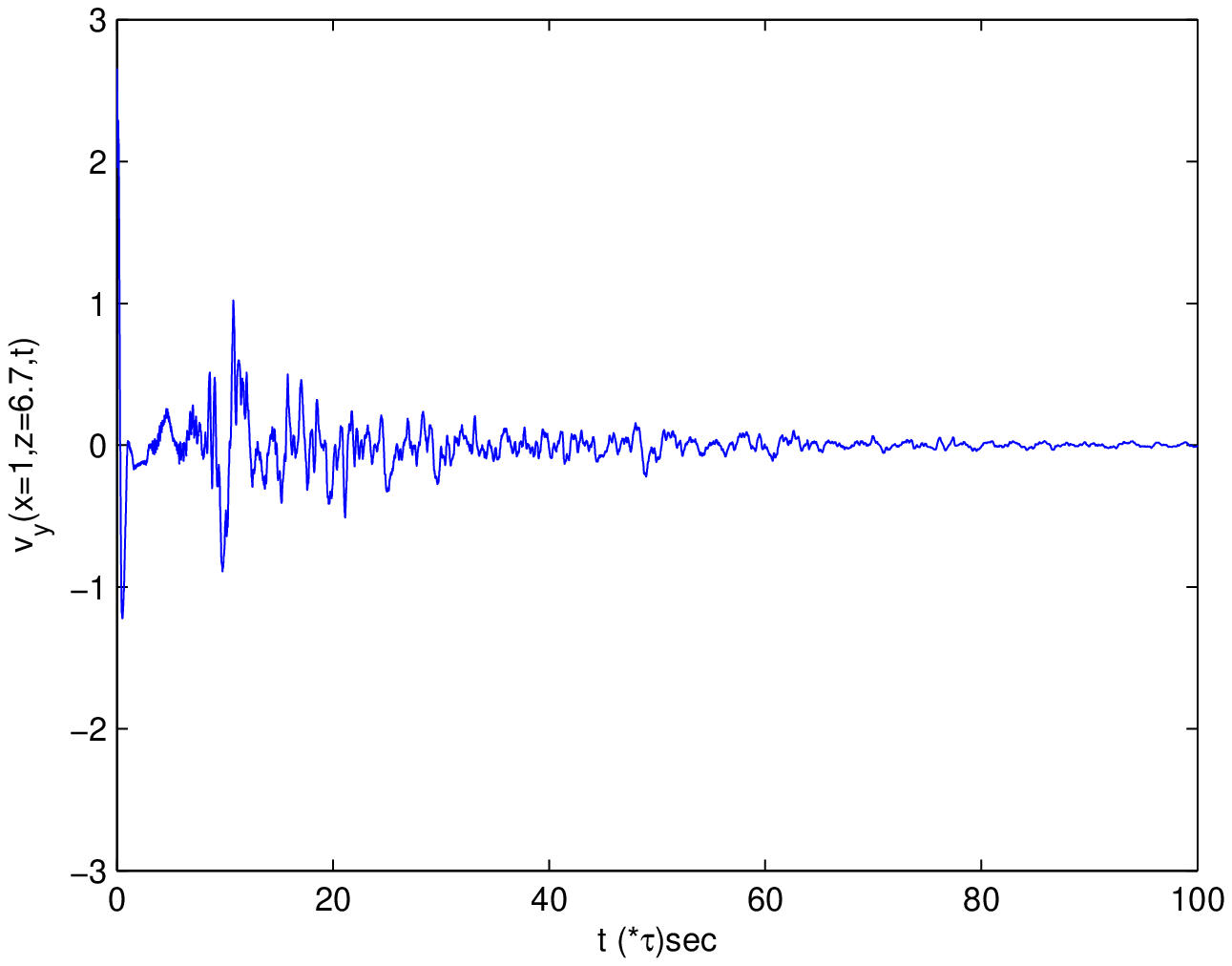}
\caption{The perturbed velocity variations with respect to time
in $x=1000$ km, $z=1300$ km; $x=1000$ km, $z=4000$ km; and $x=1000$ km,
$z=6700$ km respectively from top to bottom are showed.\label{fig2}}
\end{figure}

\begin{figure}
\centering
\includegraphics[width=8cm]{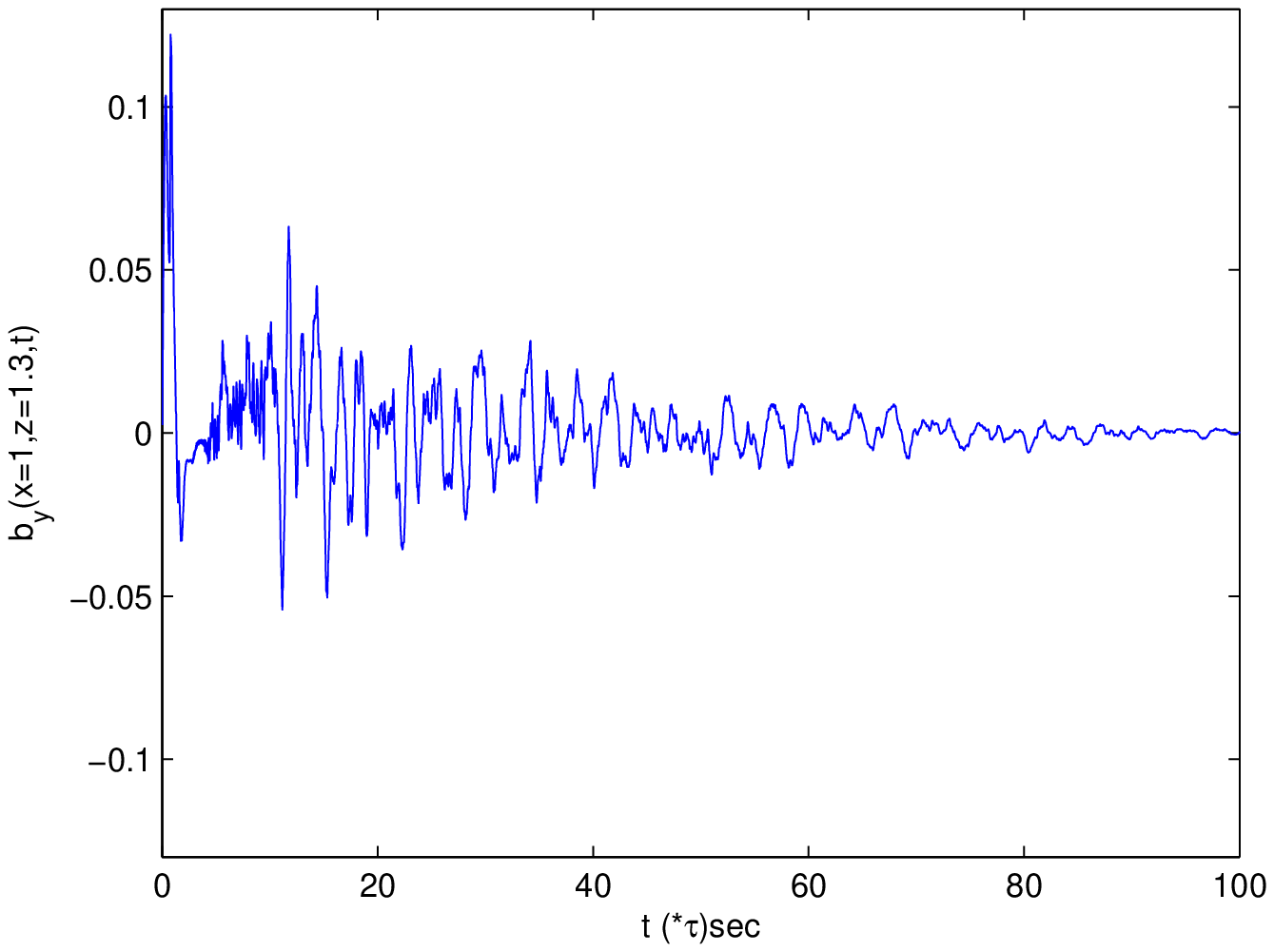}
\includegraphics[width=8cm]{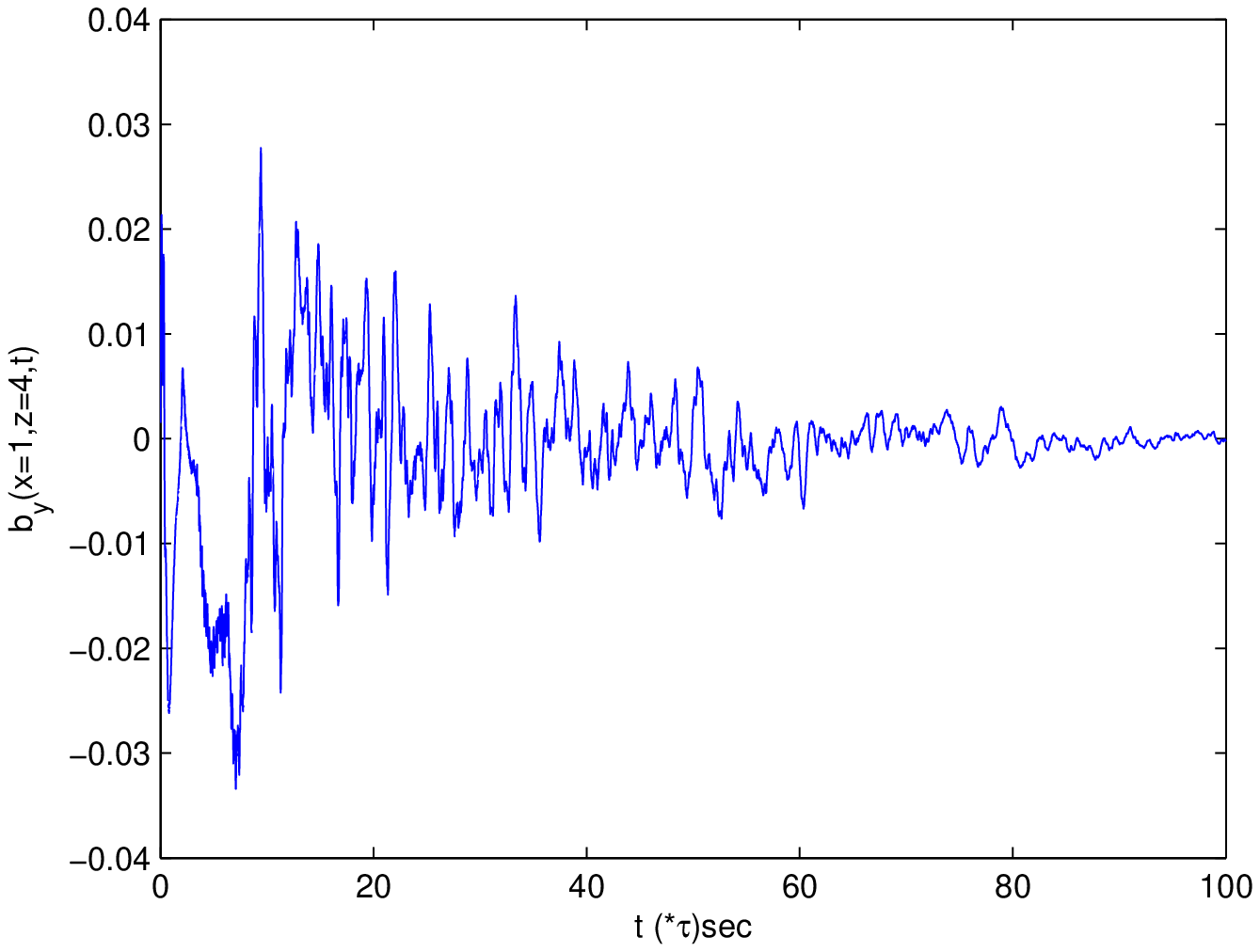}
\includegraphics[width=8cm]{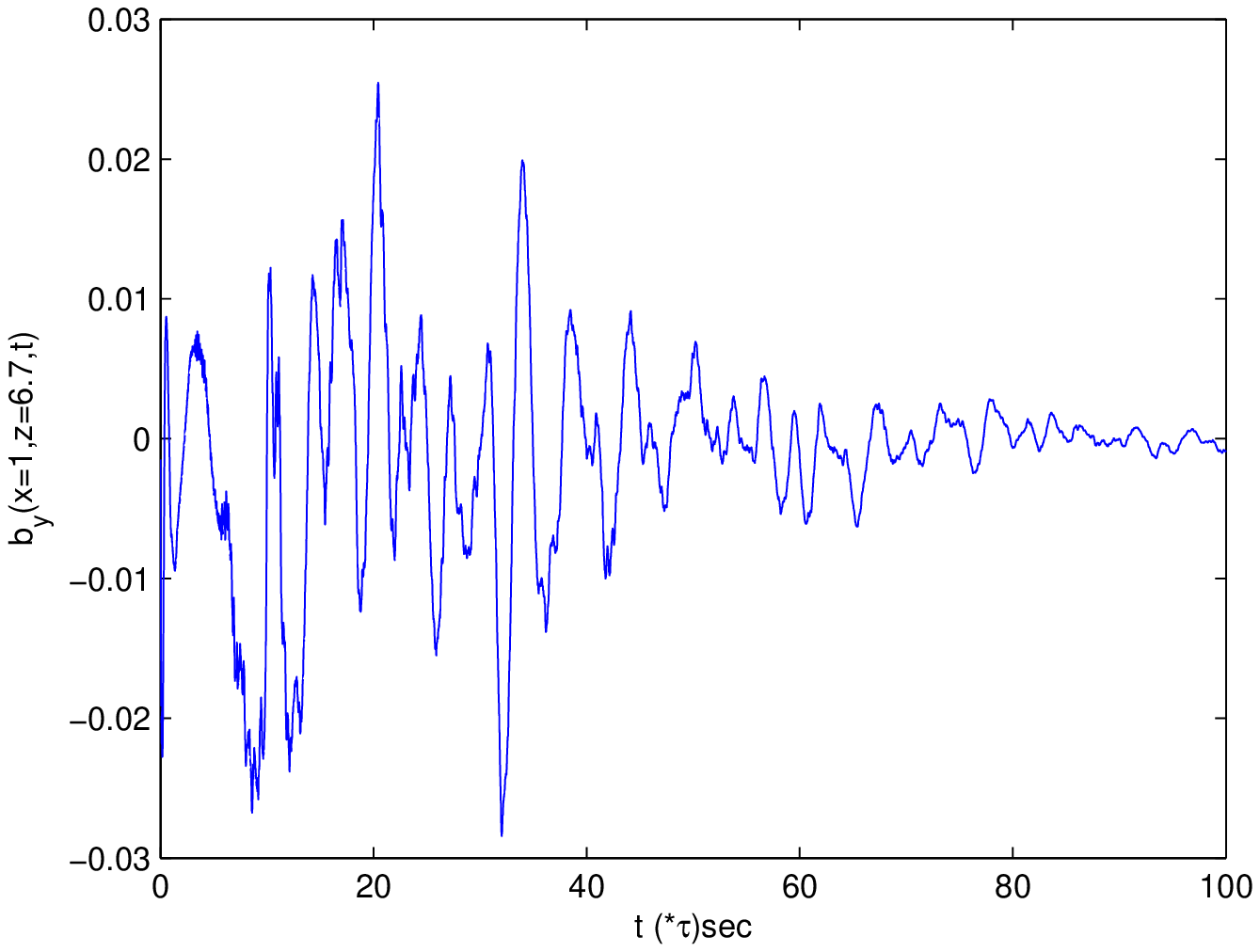}
\caption{The same as in Figure~\ref{fig2} but for the perturbed magnetic field.\label{fig3}}
\end{figure}

To solve the coupled Eqs.~\ref{eq:velo}, and~\ref{eq:mag} numerically,
the finite difference and the Fourth-Order Runge-Kutta methods
are used to take the space and time derivatives, respectively.
We set the number of mesh grid points as~$256\times256$.
In addition, the time step is chosen as $0.001$, and the system length in the $x$ and $z$ dimensions
(simulation box sizes) are set to be ($0$,$2$) and ($0$,$8$).
The parameters in spicule environment are as follows \citep{Tem2010,Ebadi2012a}:
$a$ (spicule radius)=1000 km,
$d=0.3a=300 km$ (the width of gaussian paket), L=8000 km (Spicule length), $v_{0}=25 km/s$, $B_{0}=10 G$, $n_{e}=10^{11} cm^{-3}$,
 $T=8000 K$, $g=272 m s^{-2}$, $R=8300 m^{2}s^{-1}k^{-1}$ (universal gas constant),
$V_{A0}=50 km/s$,  $k=pi/8$ (dimensionless wavenumber normalized to $a$), $k_{B}=pi/16$, $\nu=10^{3} m^{2}s^{-1}$, $\eta=10^{3} m^{2}s^{-1}$,
 $\mu=0.6$,  $H=500 km$, $\tau=20$ s, and $\alpha=2$ \citep{Okamoto2011}.\\
Figure~\ref{fig2} shows the perturbed velocity variations with respect to time
for $x=1000$ km, $z=1300$ km; $x=1000$ km, $z=4000$ km; and $x=1000$ km, $z=6700$ km respectively. We presented
the perturbed magnetic field variations, obtained from our numerical analysis, in Figure~\ref{fig3}
for $x=1000$ km, $z=1300$ km; $x=1000$ km, $z=4000$ km; and $x=1000$ km, $z=6700$ km respectively.
In these plots the perturbed velocity and
magnetic field are normalized to $V_{A0}$ and $B_{0}$ respectively.
In each set of plots it is appeared that both the perturbed
velocity and magnetic field are damped at the middle stage of phase mixing.
As height increases, the perturbed velocity amplitude does increase in contrast to the behavior of
perturbed magnetic field. This means that with an increase in height,
amplitude of velocity oscillations is expanded due to significant decrease
in density, which acts as inertia against oscillations. Similar results are observed
by time-distance analysis of Solar spicule oscillations \citep{Ebadi2012a}.
It is worth to note that the density stratification influence on the magnetic field is negligible,
which is in agreement with Solar Optical Telescope observations of Solar spicules \citep{Verth2011}.

\begin{figure}
\centering
\includegraphics[width=8cm]{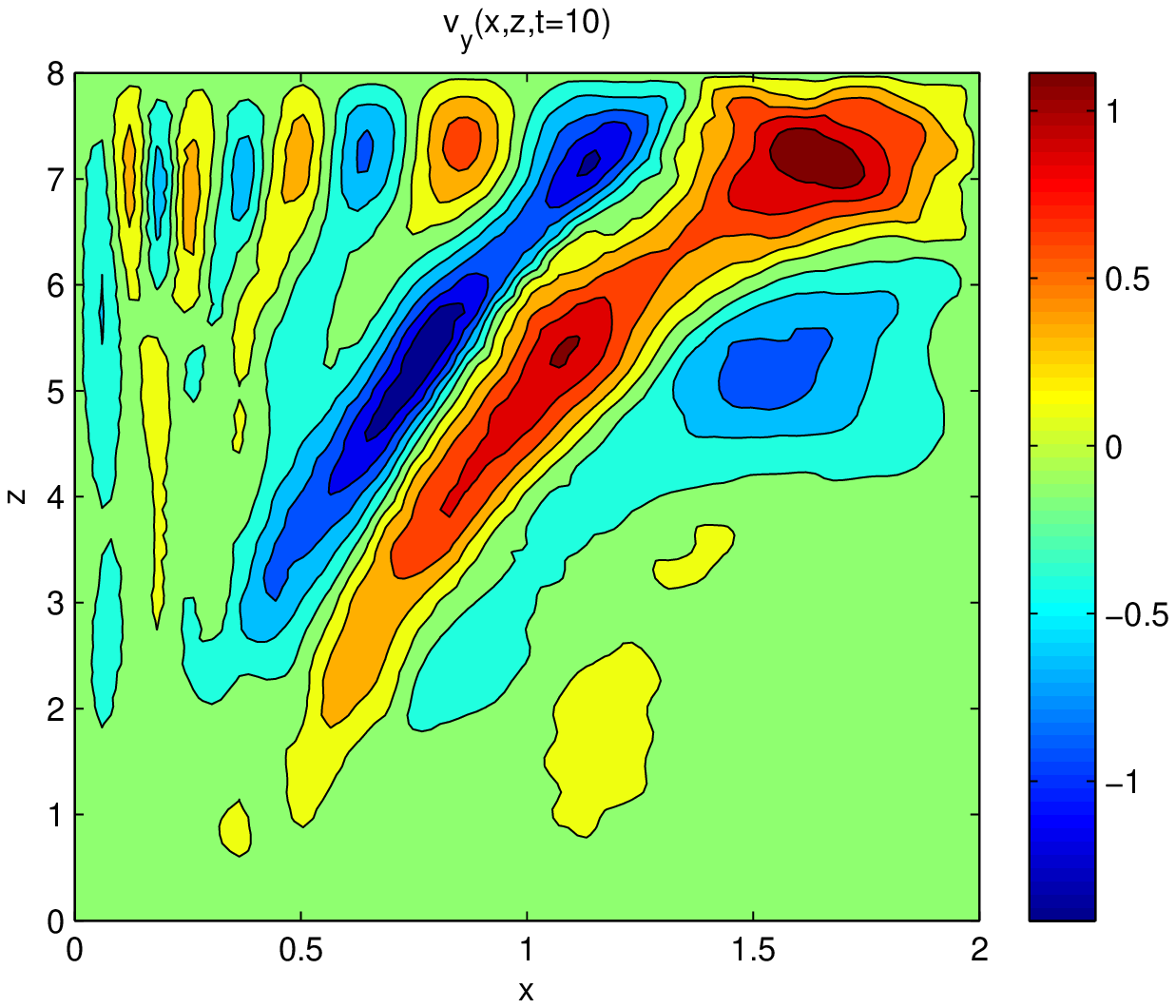}
\includegraphics[width=8cm]{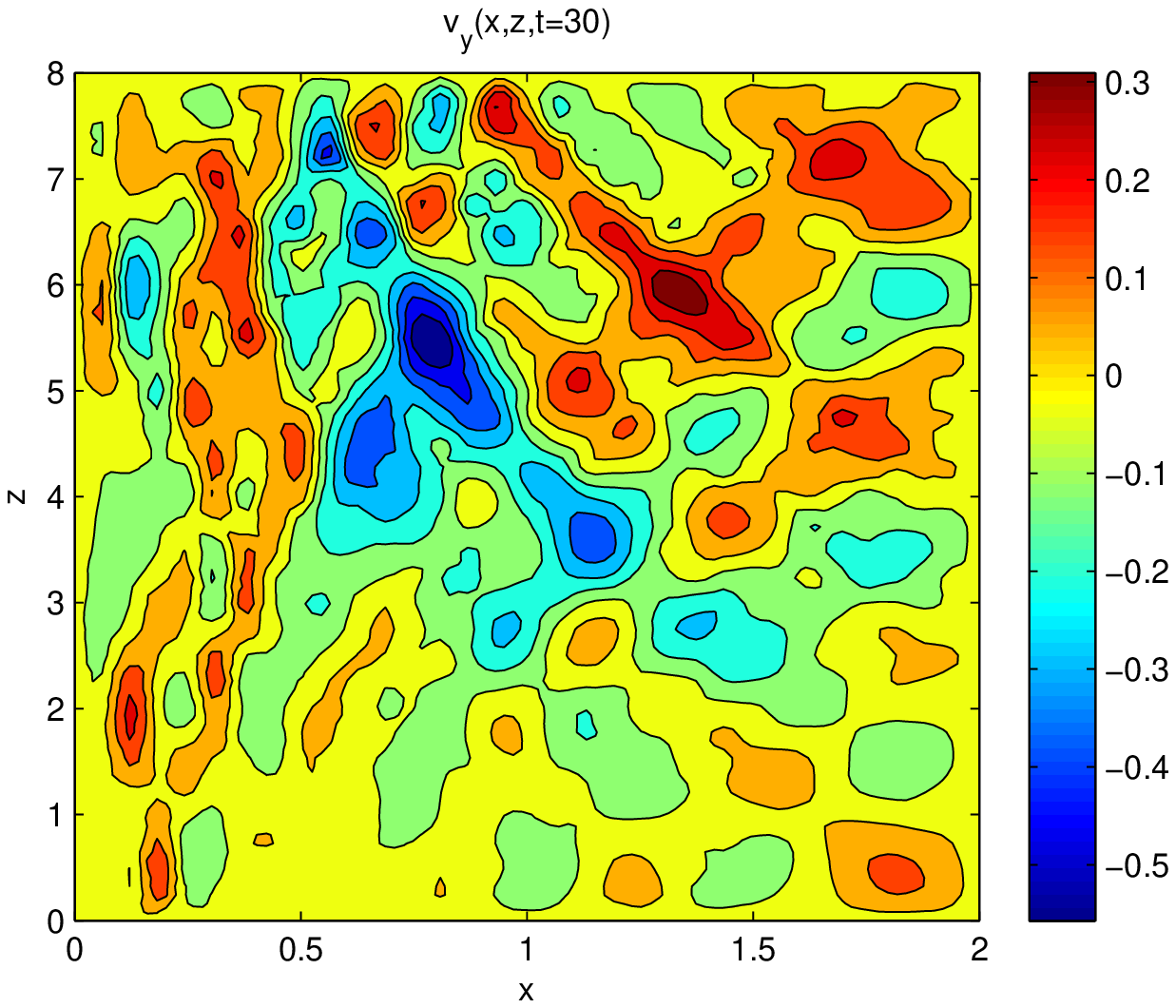}
\includegraphics[width=8cm]{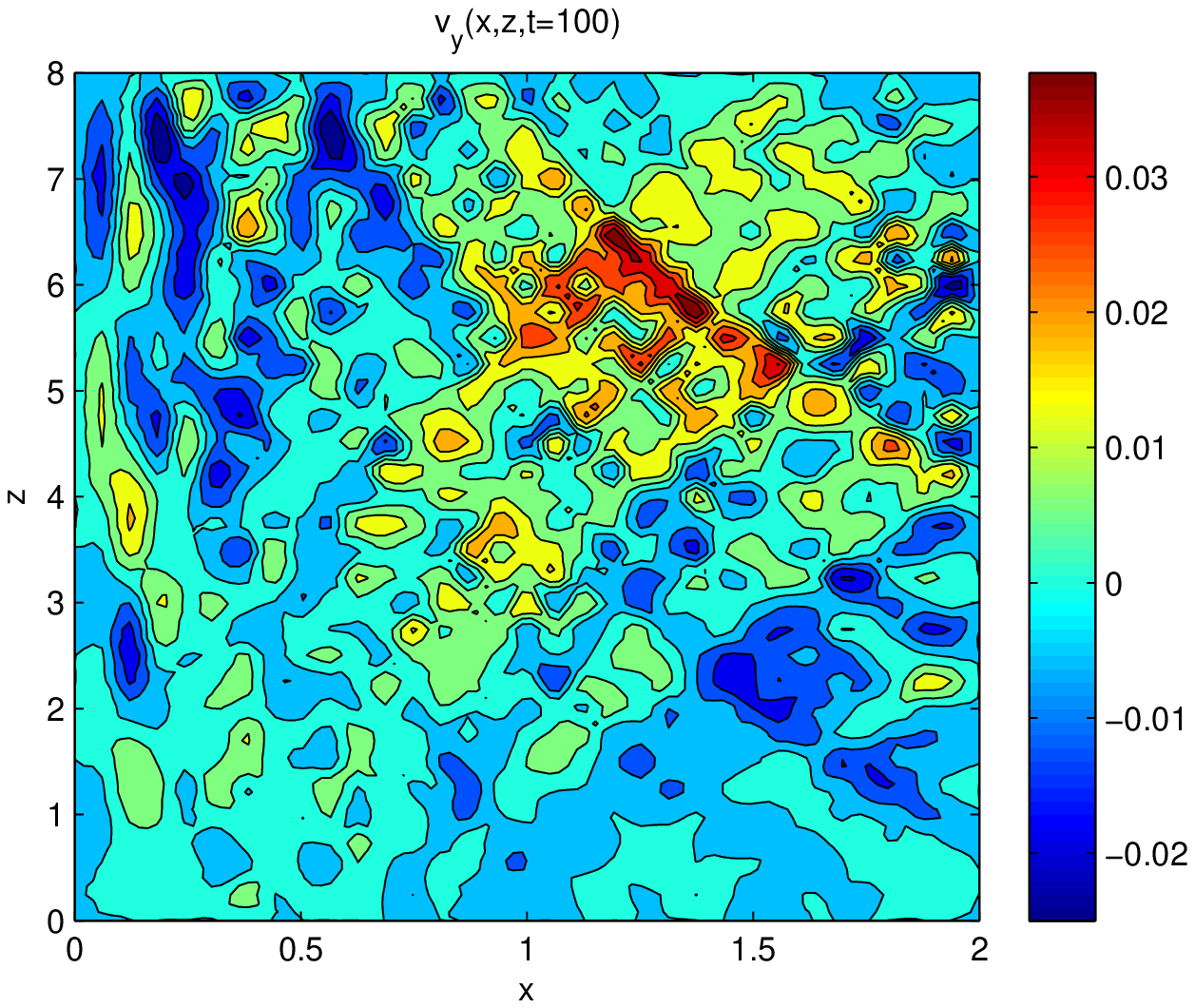}
\caption{(Color online) The perturbed velocity in $x-z$ space is presented. The panels from top to bottom correspond
to $t=10 \tau$, $t=30 \tau$, and $t=100 \tau$, respectively.    \label{fig4}}
\end{figure}

\begin{figure}
\centering
\includegraphics[width=8cm]{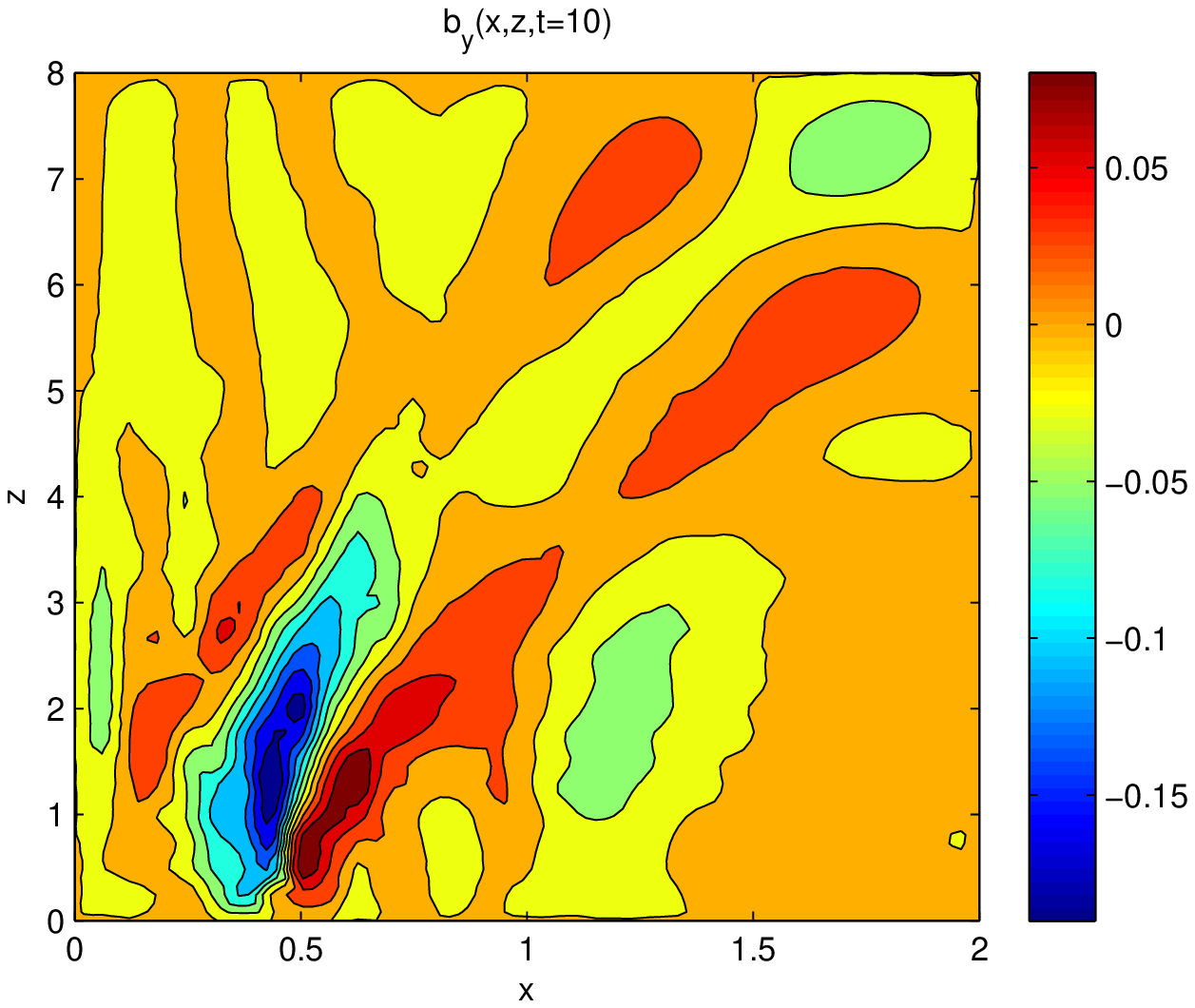}
\includegraphics[width=8cm]{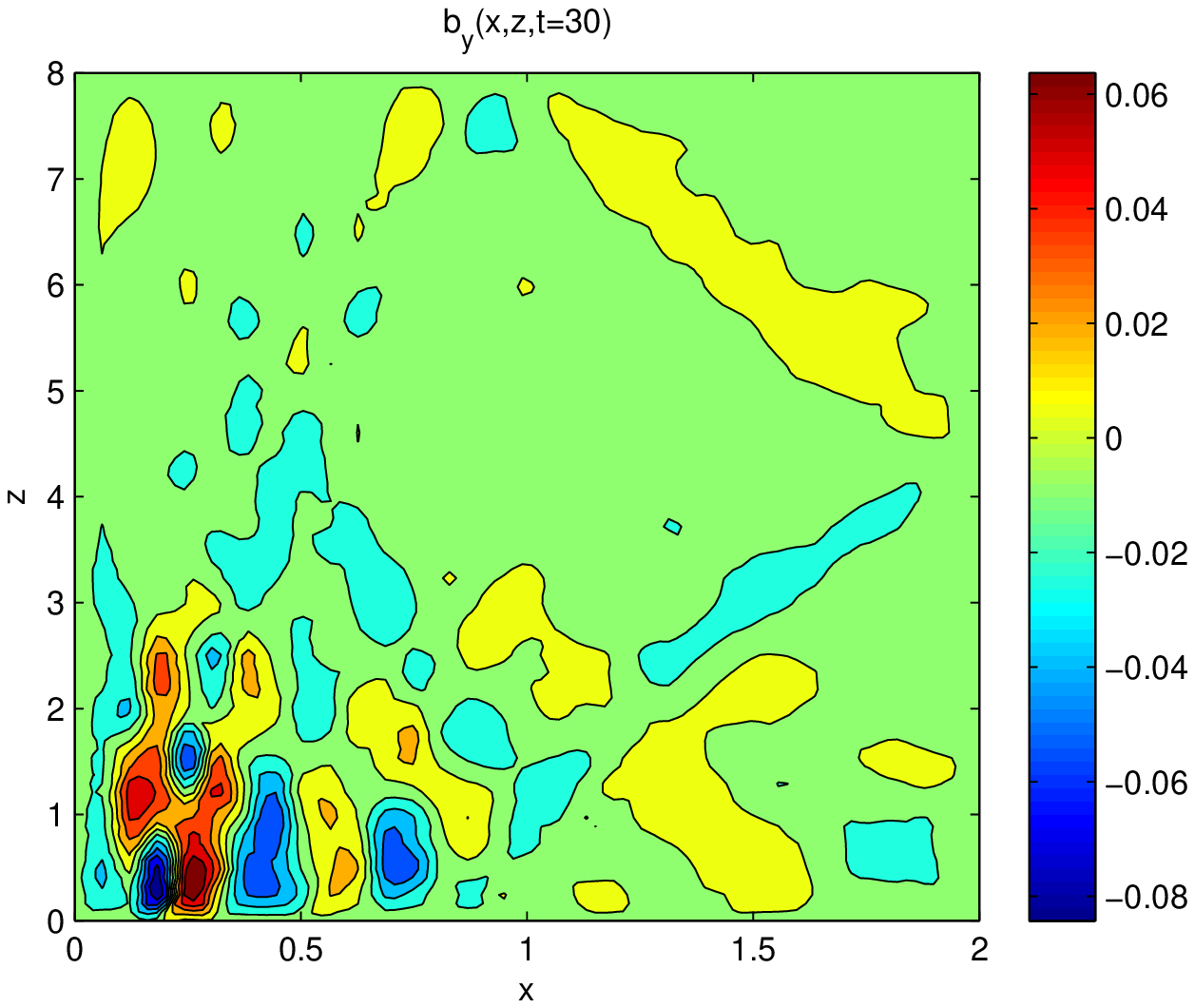}
\includegraphics[width=8cm]{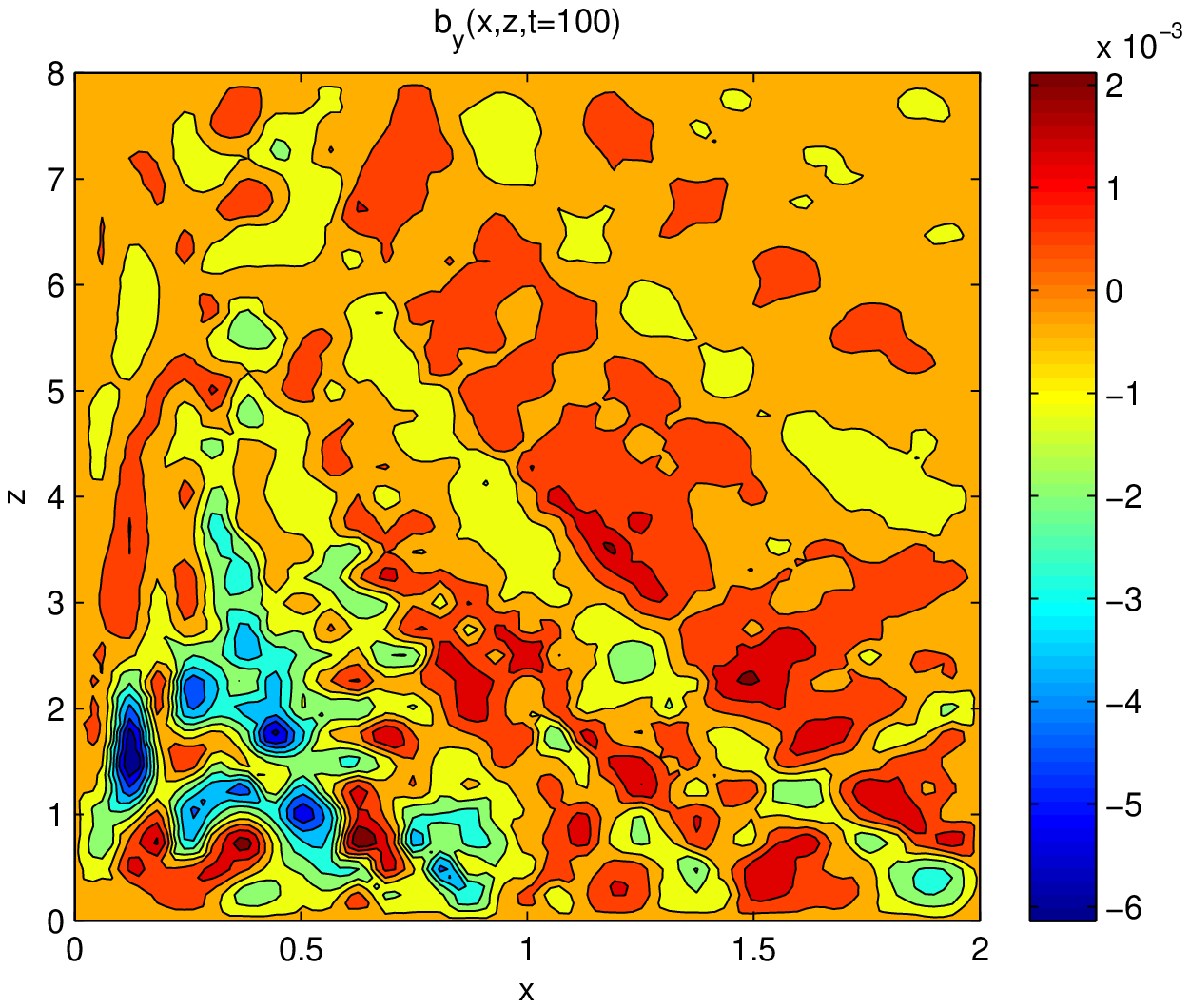}
\caption{(Color online) The same as in Figure~\ref{fig4} but for the perturbed magnetic field.\label{fig5}}
\end{figure}

Figures~\ref{fig4},~\ref{fig5} show the contour plots of the perturbed velocity and magnetic field with respect to
$x$, $z$ for $t=10 \tau$, $t=30 \tau$, and $t=100 \tau$. They show that in the presence of stratification due to gravity,
and shear flow and field the damping takes place in time as an important point of these graphs.
In other words, in spite of propagating waves, standing waves
are dissipated after a few periods due to phase mixing in the presence of shear flow and field.
Once again, increment (decrement) of the perturbed velocity (magnetic) amplitude
with height at different times are obvious from these figures. On top of these,
latter graphs show stochastic evolution pattern regarding the x-direction.
In fact, taking a package of Alfv\'{e}nic waves with different speeds at different
'x' points results in such non-uniform distribution of oscillations.
To be more rigorous, consider Eqs.~\ref{eq:velo}, and~\ref{eq:mag} where the z-derivatives are
omitted and main terms are kept only. For instance, one gets:
\begin{equation}
\label{eq:xv}
 \frac{\partial^{2} v_{y}}{\partial t^{2}}=V_{A}^{2}(x,z)\left[B_{0x}(x,z)\frac{\partial B_{0x}}{\partial x}
 \frac{\partial v_{y}}{\partial x}+B_{0x}^{2}\frac{\partial^{2} v_{y}}{\partial x^{2}}\right],
\end{equation}

for the perturbed velocity field. Differential Equation~\ref{eq:xv}, simply, implies that $v_{y}$ depends
to 'x' but unfortunately not in the form of simple function.
Additionally, having ignored the z-derivatives, still, $v_{y}$ varies with 'z' through $V_{A}(x,z)$ and $B_{0x}(x,z)$ terms.
Note that, although, one can discuss about the above point in detail, but, here, we are more
interested in studying the effects of phase mixing on time evolution and propagation
of Alfv\'{e}nic waves in the propagation direction rather than looking for variations in the specific perpendicular 'x' direction.


To estimate the damping time it is suitable to calculate the total energy (kinetic energy plus magnetic energy)
per unit of length in $y$ direction as:

\begin{equation}
\label{eq:tenergy}
 E_{tot}(t) = \frac{16\pi}{B_{0}^{2}aL} \int_{0}^{2} dx \int_{0}^{8} dz [\rho (x,z) v_{y}^{2}(x,z) + b_{y}^{2}(x,z)].
\end{equation}

\begin{figure}
\centering
\includegraphics[width=8cm]{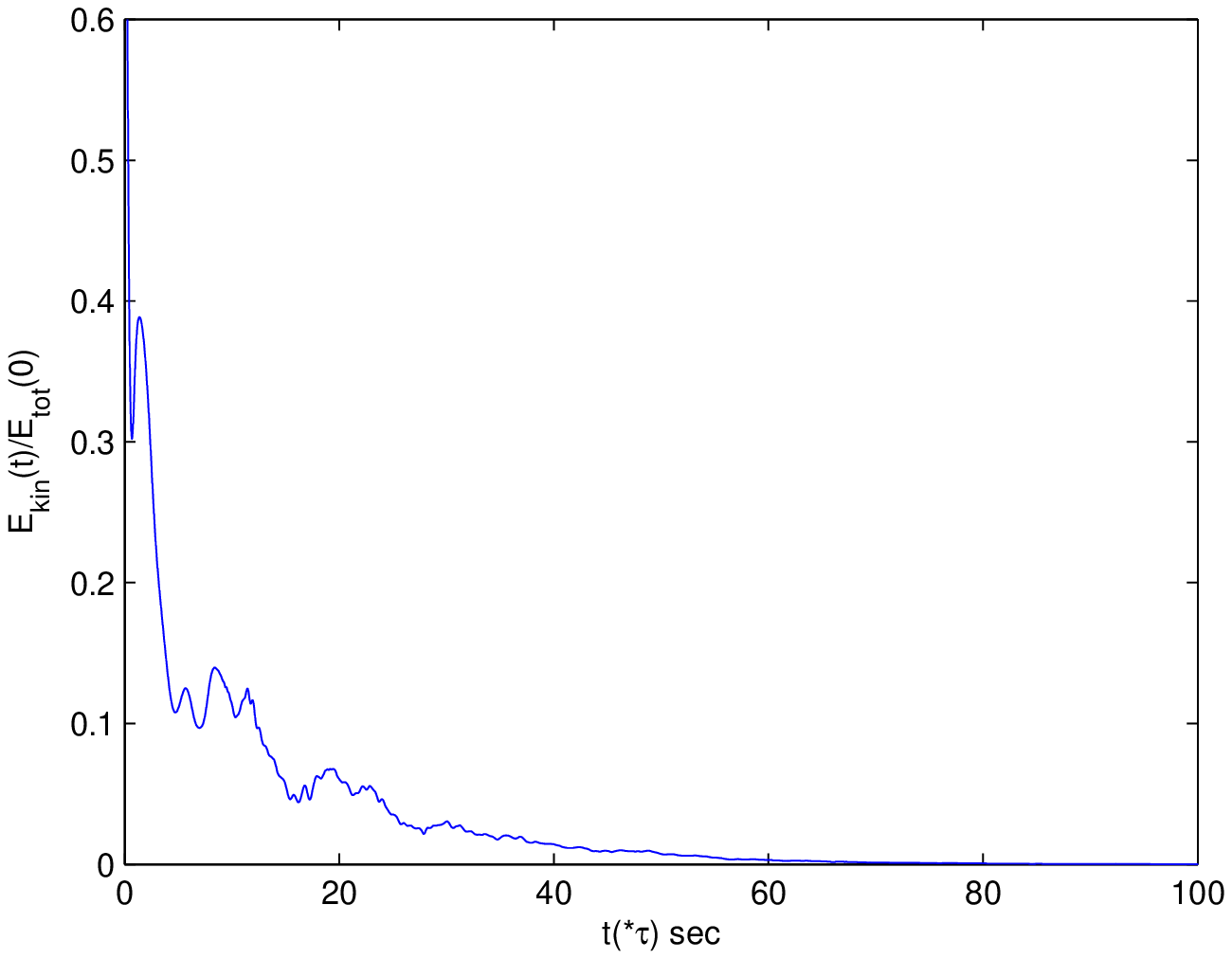}
\includegraphics[width=8cm]{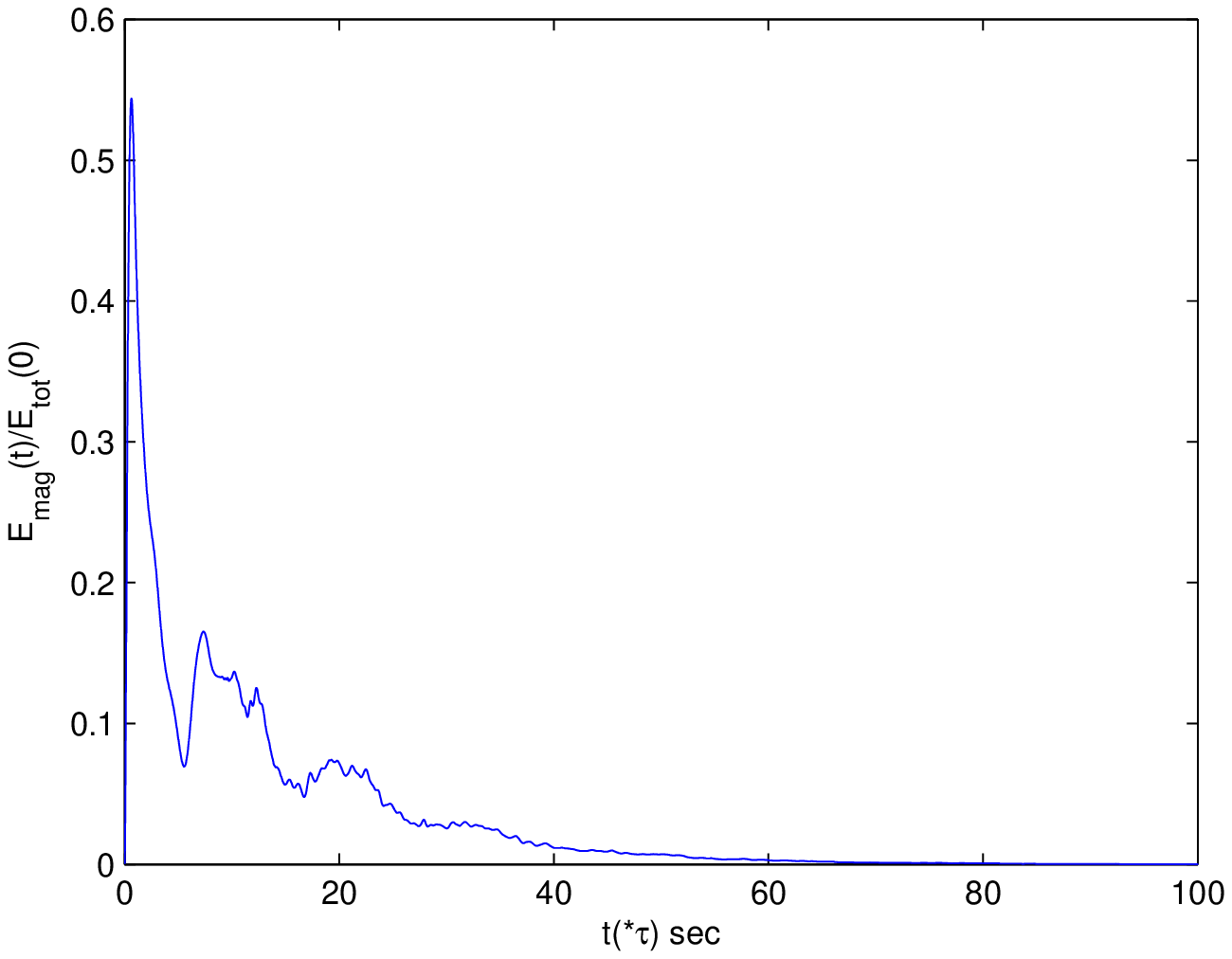}
\includegraphics[width=8cm]{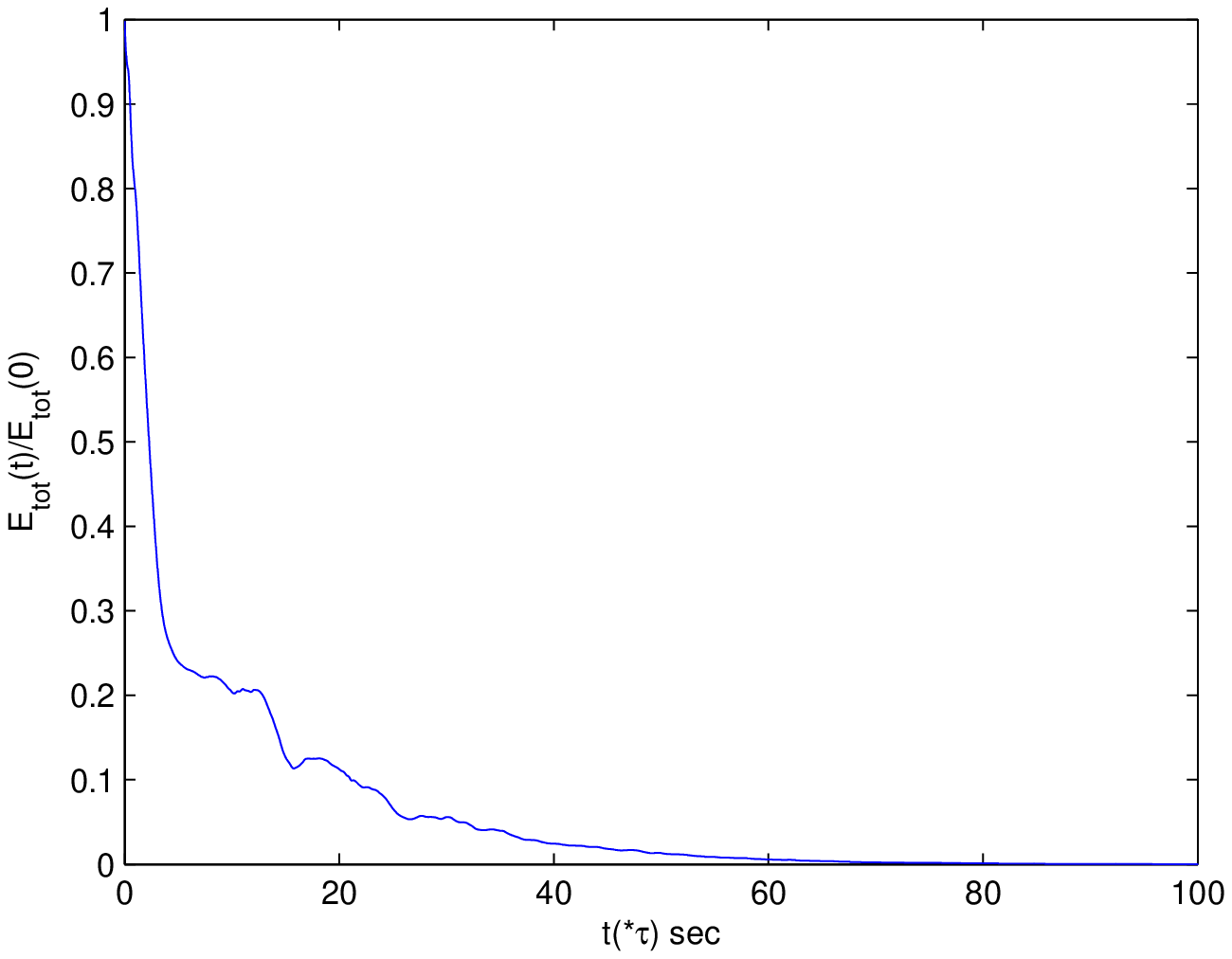}
\caption{Time variations of normalized kinetic energy, magnetic energy, and total energy for $d=0.3a$
are presented from top to bottom respectively.\label{fig6}}
\end{figure}

In Figure~\ref{fig6} the kinetic energy, magnetic energy, and total energy
normalized to the initial total energy are presented respectively from top to bottom.
It is clear from energy plots that the standing Alfv\'{e}n waves under spicule conditions
can dissipated after a few periods. This is in agreement with the fact that spicules have short lifetimes,
and are transient phenomena. The energy flux, stored in oscillating spicule axis, is of the order of coronal energy
loss in quiet sun. Therefore, dissipation of standing Alfv\'{e}n waves can count as a candidate mechanism in transferring
and as a result heating the solar corona. Moreover, it should be emphasized that the stratification due to gravity, shear flow
and field are the facts that should exist in MHD models in spicules.

\section{Conclusion}
\label{sec:concl}
In our model, we assume that spicules are small scale structures
with an initial shear flow and field, and a uniform temperature along them.
Density variation along the spicule axis is considerable, and stratification due
to gravity is significant. As a result, the medium is dense in its lower heights,
but it becomes rare and rare as height increases.
The perturbed velocity amplitude does increase in contrast to the behavior of
perturbed magnetic field. This means that with an increase in height,
amplitude of velocity oscillations is expanded due to stratification.
It is worth to note that the density stratification influence on the magnetic field is negligible.
The divergent configuration of initial magnetic field with
sheared plasma flow can fasten the phase mixing and dissipation of standing Alfv\'{e}n waves in Spicules.
This is in agreement with the fact that spicules have short lifetimes, and are disappeared after a few periods.

\acknowledgments
This work has been supported financially by the Research Institute for Astronomy and Astrophysics of Maragha (RIAAM), Maragha, Iran.

\makeatletter
\let\clear@thebibliography@page=\relax
\makeatother

\end{document}